\documentclass[preprint]{article}
\usepackage{authblk}
\usepackage{graphicx}
\usepackage{dcolumn}
\usepackage{bm}
\usepackage{amsmath}
\usepackage{amssymb}
\usepackage{multirow}
\usepackage{caption}
\usepackage{subcaption}
\usepackage{epstopdf}
\usepackage{hyperref}
\usepackage{cite}
\begin{document}
\title{\bf A new parametric observational study of $f(Q,B)$ gravity with modified chaplygin gas }
\author[]{Amit Samaddar\thanks{samaddaramit4@gmail.com}}
\author[]{S. Surendra Singh\thanks{ssuren.mu@gmail.com}}
\affil[]{Department of Mathematics, National Institute of Technology Manipur, Imphal-795004,India.}

\maketitle

\textbf{Abstract}: In this work, we explore the cosmological dynamics of a modified gravity framework based on the function $f(Q,B)=\delta Q^{2}+\beta B$, where $Q$ denotes the nonmetricity scalar and $B$ is the boundary term that relates $Q$ to the Ricci scalar. The matter sector is modeled using the Modified Chaplygin Gas (MCG) with the equation of state $p=A\rho-\frac{B}{\rho^{\alpha}}$, allowing the model to interpolate between early-time matter behavior and late-time cosmic acceleration. By deriving an analytical expression for the Hubble parameter $H(z)$, we perform a parameter estimation using Markov Chain Monte Carlo (MCMC) techniques in conjunction with the latest cosmological observations: $46$ Hubble parameter measurements, $15$ BAO data points, DESI DR2 BAO data and the Pantheon+ Type Ia supernovae compilation. The best-fit values are obtained as $H_0 = 72.22^{+3.64}_{-4.46}$, $A_s = 0.696^{+0.082}_{-0.129}$, $\alpha = 0.0029^{+0.022}_{-0.021}$, and $A = 0.0038^{+0.071}_{-0.047}$. The deceleration parameter transitions at redshift $z_{tr} \approx 0.946$, while the present-day value is $q_0 = -0.789$. The model yields an age of the Universe $t_0 \approx 13.53$ Gyr and a present EoS parameter $\omega_0 \approx -0.691$, which reflects the late-time acceleration consistent with observational bounds. These results demonstrate that the MCG scenario within $f(Q,B)$ gravity provides a viable and observationally consistent framework for explaining the late-time accelerated expansion of the Universe.


\section{Introduction}\label{sec1}
\hspace{0.5cm} The Universe has experienced multiple evolutionary stages after the Big Bang—from an initial radiation-dominated phase to matter domination and eventually transitioning into an accelerated expansion phase, first revealed by Type Ia supernovae observations \cite{Riess98}. This late-time acceleration, now strongly supported by various cosmological probes such as cosmic microwave background (CMB) measurements \cite{Planck16}, baryon acoustic oscillations (BAO) \cite{Cole05} and large-scale structure surveys \cite{Tde15}, remains one of the most profound mysteries in modern cosmology. The standard $\Lambda$CDM model attributes this accelerated expansion to a cosmological constant $(\Lambda)$, interpreted as dark energy with constant equation of state $(\omega=-1)$. Despite its observational success, $\Lambda$CDM faces persistent theoretical problems, including the fine-tuning and coincidence issues \cite{Swe89,Pad03,Sahni06}, as well as emerging tensions in key parameters like the Hubble constant $H_{0}$ between early and late-time measurements \cite{ED21}. Such challenges have driven interest in developing alternative gravitational theories that extend beyond Einstein’s General Relativity (GR). Specifically, modified gravity models seek to explain the observed cosmic acceleration without relying on hypothetical dark energy components. One promising direction involves geometrically extending GR through modifications involving curvature, torsion, or nonmetricity, each providing a coherent framework that reshapes the fundamental geometry of spacetime \cite{TP10,Nojiri11,Mcap11}.

A particularly intriguing and rapidly evolving area is the symmetric teleparallel formulation of gravity, in which gravitational effects arise from the nonmetricity of spacetime, rather than from curvature (as in GR) or torsion (as in teleparallel gravity). This approach gives rise to $f(Q)$ gravity, where $Q$ represents the nonmetricity scalar derived from a connection that is both flat and torsion-free \cite{JB18,Xu19,Hoh19}. In contrast to $f(R)$ or $f(T)$ theories, $f(Q)$ gravity yields second-order field equations and naturally avoids issues like ghost instabilities \cite{Mandal20,Singh23}. To increase the dynamical flexibility of $f(Q)$ gravity and allow for a smoother recovery of GR, the theory can be extended by incorporating boundary terms. This leads to the generalized $f(Q, B)$ framework, where $B$ is a boundary term that links the Ricci scalar $R$ from GR to the nonmetricity scalar through the relation $R=-Q+B$. This generalization, recently introduced in the literature \cite{Capone21,Harko18}, expands the range of possible cosmological solutions while preserving the second-order nature of the field equations. Notably, in the coincident gauge—where the affine connection vanishes—the resulting field equations become especially simple, aiding both in theoretical exploration and numerical computations. The $f(Q, B)$ gravity framework belongs to the same geometric family as other fourth-order theories like $f(R)$, $f(T, B)$, and scalar–nonmetricity models \cite{Tharko11,Lazkoz17,Ranjit22,Amit23}, all of which aim to extend the geometric foundations of gravity to explain cosmic evolution without invoking a cosmological constant. Various extensions—such as $f(Q,\mathcal{L}_m)$ \cite{Ataz12,A2025}, $f(Q, T)$ \cite{SB22}, and models incorporating scalar fields \cite{Sola22,JBJ20}—have been developed to tackle both early- and late-time cosmological phenomena, including inflation, dark energy and the unification of dark matter and dark energy.

Among the many phenomenological models developed to explain the dark sector of the Universe, Chaplygin gas models have garnered significant attention due to their potential to unify dark matter and dark energy within a single fluid description. Initially derived from principles in aerodynamics, the Chaplygin gas EoS was formulated by S. Chaplygin in $1904$ \cite{Chap04}. Its relevance to cosmology emerged much later, when it was discovered that this type of fluid naturally evolves from a matter-dominated (dust-like) behavior in the early universe to a cosmological constant–like behavior at late times. Despite this appealing feature, the original Chaplygin gas model faced difficulties, particularly in explaining structure formation and matching CMB anisotropy observations, which led to the development of various generalizations. The Generalized Chaplygin Gas (GCG) model \cite{Bento2002} improved upon the original equation of state by introducing an additional parameter, offering greater flexibility in the evolution of the cosmic fluid and enhancing agreement with observational data. To further develop the model and accommodate radiation-like behavior at high energy densities, the Modified Chaplygin Gas (MCG) was introduced \cite{Benaoum2002,Jamil11,Debnath11}. This version added a linear barotropic term, enriching the dynamical properties of the model. As a result, the MCG can effectively emulate radiation, matter, and dark energy phases of the universe depending on the chosen parameters, positioning it as a strong candidate for a unified description of the cosmic dark sector. The Modified Chaplygin Gas (MCG) model has been extensively studied in the context of observational cosmology. For example, \cite{Debnath21} explored the cosmological implications of Chaplygin-Jacobi and Chaplygin-Abel gases within the FRW framework. Paul and Thakur \cite{Paul13} analyzed MCG constraints using cosmic growth rate measurements, while Lu et al. \cite{Lu08} derived parameter bounds based on CMB, BAO, and supernova datasets, evaluating the model’s performance relative to other dark energy scenarios. Additionally, Paul and Manna \cite{Paul17} investigated observational limits on extended Chaplygin gas models, emphasizing the adaptability of the framework when tested against diverse observational datasets. Motivated by these successes, in this work, we extend the analysis of the MCG model to the framework of $f(Q,B)$ gravity, an extension of symmetric teleparallel gravity involving the non-metricity scalar $Q$ and its boundary term $B$. By embedding the MCG within this modified geometrical context, we investigate its cosmological implications and confront it with the latest observational datasets, aiming to understand how the geometric structure of gravity influences the dynamics of unified dark fluids.

To evaluate the consistency of our model with observations, we perform a joint analysis using the latest cosmological data. This includes $46$ measurements of the Hubble parameter, $15$ baryon acoustic oscillation (BAO) data points, the most recent DESI DR2 BAO sample, and the Pantheon+ dataset comprising $1701$ Type Ia supernovae. This comprehensive approach enables us to tightly constrain the model parameters and assess its compatibility with present-day cosmological observations.

The structure of the paper is as follows: In section \ref{sec2}, we present the field equations of $f(Q,B)$ gravity along with the chosen non-linear model. Section \ref{sec3} is devoted to the MCG, where we derive the corresponding $H(z)$ expression for our model. In section \ref{sec4}, we describe the observational datasets used and employ them to constrain the model parameters. Section \ref{sec5} focuses on the analysis of key cosmological parameters derived from the best-fit values. The energy conditions are examined in section \ref{sec6}, followed by a dynamical study in the $\omega$–$\omega'$ plane in section \ref{sec7}. The age of the Universe is calculated in section \ref{sec8}. Finally, we summarize our findings and draw conclusions in section \ref{sec9}.
\section{Field equations and cosmological dynamics in $f(Q,B)$ gravity theory}\label{sec2}
\hspace{0.5cm} Modified gravity theories that utilize the nonmetricity tensor have emerged as promising alternatives to General Relativity (GR). Unlike GR, where gravity arises from the curvature of spacetime, these theories attribute gravitational effects to nonmetricity. In this context, the underlying connection is specifically constructed to eliminate both curvature and torsion, resulting in what is known as symmetric teleparallel geometry. This geometric formulation is characterized by the following defining properties: $(i)$ Curvature tensor vanishes: $R^{\epsilon}_{\mu\nu}=0$, $(ii)$ Torsion tensor vanishes: $T^{\epsilon}_{\mu\nu}=0$ and $(iii)$ Nonmetricity is non-zero: $Q_{\lambda\mu\nu}=\nabla_{\lambda}g_{\mu\nu}\neq0$. These conditions are satisfied by adopting the coincident gauge, in which the affine connection $\Gamma_{\mu\nu}^{\lambda}$ is set to vanish globally \cite{YNO21}:
\begin{equation}\label{1}
\Gamma_{\mu\nu}^{\lambda}=0,
\end{equation}
This gauge simplifies the theory by eliminating inertial effects and ensuring that the covariant derivative reduces to partial derivatives. 

The fundamental geometric object in symmetric teleparallel gravity is the nonmetricity tensor, defined as:
\begin{equation}\label{2}
Q_{\lambda\mu\nu}\equiv\nabla_{\lambda}g_{\mu\nu}=\partial_{\lambda}g_{\mu\nu}-\Gamma^{\epsilon}_{\lambda\mu}g_{\epsilon\nu}-\Gamma^{\epsilon}_{\lambda\nu}g_{\mu\epsilon},
\end{equation}
From this, we construct two traces:
\begin{equation}\label{3}
Q_{\lambda}=Q_{\lambda}{}^{\mu}{}_{\mu}, \hspace{0.7cm} \widetilde{Q}_{\lambda}=Q^{\mu}{}_{\lambda\mu}.
\end{equation}
The superpotential tensor $P^{\lambda}_{\mu\nu}$ is defined to ensure the correct dynamics from the action:
\begin{equation}\label{4}
P^{\lambda}{}_{\mu\nu} = \frac{1}{4} \left[- Q^{\lambda}{}_{\mu\nu}+ 2 Q_{(\mu}{}^{\lambda}{}_{\nu)}- Q^{\lambda} g_{\mu\nu}+ \bar{Q}^{\lambda} g_{\mu\nu}- \delta^{\lambda}_{(\mu} Q_{\nu)}\right]
\end{equation}
which is symmetric in the lower two indices due to the symmetry of the connection.

Using this, the nonmetricity scalar is constructed as:
\begin{equation}\label{5}
Q=-Q_{\lambda\mu\nu}P^{\lambda\mu\nu},
\end{equation}
In the standard Levi-Civita connection, which is metric-compatible, the Ricci scalar $R$ arises purely from the spacetime curvature. However, in the framework of symmetric teleparallel gravity, the scalar $R$ is connected to the nonmetricity scalar $Q$ through an additional boundary term $B$. This relationship can be expressed as:
\begin{equation}\label{6}
R=-Q+B,
\end{equation}
where
\begin{equation}\label{7}
B=\frac{2}{\sqrt{-g}}\partial_{\mu}(\sqrt{-g}Q^{\mu}),
\end{equation}
This relation enables General Relativity to be re-expressed using the nonmetricity scalar $Q$ and the boundary term $B$, where $B$ has no dynamical influence in conventional GR. However, this equivalence no longer holds if the gravitational action involves nonlinear functions of $Q$ and $B$.

We generalize the gravitational action by promoting the Ricci scalar replacement $R=-Q+B$ into an arbitrary function $f(Q,B)$. The total action, including matter contributions, becomes \cite{Falco23}:
\begin{equation}\label{8}
S=\int d^{4}x\sqrt{-g}\bigg[\frac{1}{2\kappa}f(Q,B)+\mathcal{L}_{m}\bigg],
\end{equation}
The field equations are obtained by performing a variation of the action with respect to the metric tensor.
\begin{eqnarray}\label{9}
\kappa T_{\mu\nu}&=&-\frac{f}{2}g_{\mu\nu}+\frac{2}{\sqrt{-g}}\partial_{\lambda}\bigg(\sqrt{-g}f_{Q}P^{\lambda}{}_{\mu\nu}\bigg)+\bigg[P_{\mu\sigma\delta}Q_{\nu}{}^{\sigma\delta}
-2P_{\sigma\delta\nu}Q^{\sigma\delta}{}_{\mu}\bigg]f_{Q}\\\nonumber
&&+\bigg(\frac{B}{2}g_{\mu\nu}-\overset{\circ}\nabla_{\mu}\overset{\circ}\nabla_{\nu}+g_{\mu\nu}\overset{\circ}\nabla{}^{\sigma}\overset{\circ}\nabla_{\sigma}-2P^{\lambda}{}_{\mu\nu}\partial_{\lambda}
\bigg)f_{B}.
\end{eqnarray}
In a fully covariant form, the field equations of $f(Q, B)$ gravity can be reformulated to make the geometric contributions explicit, where we denote $f_Q = \frac{\partial f}{\partial Q}$ and $f_B = \frac{\partial f}{\partial B}$ as the partial derivatives of the gravitational Lagrangian with respect to the nonmetricity scalar and the boundary term, respectively.
\begin{eqnarray}\label{10}
\kappa T_{\mu\nu}&=&-\frac{f}{2}g_{\mu\nu}+\frac{2}{\sqrt{-g}}P^{\lambda}{}_{\mu\nu}\nabla_{\lambda}(f_{Q}-f_{B})+\bigg[\overset{\circ}G_{\mu\nu}+\frac{Q}{2}g_{\mu\nu}\bigg]f_{Q}\\\nonumber
&&+\bigg(\frac{B}{2}g_{\mu\nu}-\overset{\circ}\nabla_{\mu}\overset{\circ}\nabla_{\nu}+g_{\mu\nu}\overset{\circ}\nabla{}^{\sigma}\overset{\circ}\nabla_{\sigma}\bigg)f_{B}.
\end{eqnarray}
To isolate the contributions from the geometry and express the field equations in a form analogous to Einstein’s equations, one can define an effective energy-momentum tensor that absorbs all additional terms arising from the modified geometry. Specifically, we define:
\begin{equation}\label{11}
T^{(eff)}_{\mu\nu}=T_{\mu\nu}+\frac{1}{\kappa}\bigg[\frac{f}{2}g_{\mu\nu}-2P^{\lambda}{}_{\mu\nu}\nabla_{\lambda}(f_{Q}-f_{B})-\frac{1}{2}Qf_{Q}g_{\mu\nu}
-\bigg(\frac{B}{2}g_{\mu\nu}-\overset{\circ}\nabla_{\mu}\overset{\circ}\nabla_{\nu}+g_{\mu\nu}\overset{\circ}\nabla{}^{\sigma}\overset{\circ}\nabla_{\sigma}\bigg)f_{B}\bigg],
\end{equation}
This allows the gravitational field equations to be recast in a form resembling the standard Einstein equations:
\begin{equation}\label{12}
\overset{\circ}G_{\mu\nu}=\frac{\kappa}{f_{Q}}T^{(eff)}_{\mu\nu},
\end{equation}
where $\overset{\circ}G_{\mu\nu}$ is the Einstein tensor constructed from the Levi-Civita connection. In this representation, all modifications to GR from the extended geometry are encoded into the effective stress-energy tensor.

To explore the cosmological consequences of $f(Q, B)$ gravity, we assume a spatially flat, homogeneous, and isotropic universe. This is modeled using the conventional Friedmann–Lema$\acute{i}$tre–Robertson–Walker (FLRW) metric expressed in Cartesian coordinates:
\begin{equation}\label{13}
ds^{2}=-dt^{2}+a^{2}(t)(dx^{2}+dy^{2}+dz^{2}),
\end{equation}
Here, $a(t)$ represents the scale factor of the universe, while the Hubble parameter $H(t)$ is given by the ratio $\frac{\dot{a}(t)}{a(t)}$, where the dot signifies a derivative with respect to cosmic time.

Within this spacetime background, the field equations obtained from the $f(Q, B)$ framework show that the modified geometric structure introduces extra contributions beyond the standard matter fields. These additional terms give rise to an effective geometric fluid, which can be viewed as a manifestation of dark energy. The corresponding energy-momentum tensor for this component is expressed as:
\begin{equation}\label{14}
T^{(DE)}_{\mu\nu}=\frac{1}{f_{Q}}\bigg[\frac{f}{2}g_{\mu\nu}-2P^{\lambda}{}_{\mu\nu}\nabla_{\lambda}(f_{Q}-f_{B})-\frac{1}{2}Qf_{Q}g_{\mu\nu}
-\bigg(\frac{B}{2}g_{\mu\nu}-\overset{\circ}\nabla_{\mu}\overset{\circ}\nabla_{\nu}+g_{\mu\nu}\overset{\circ}\nabla{}^{\sigma}\overset{\circ}\nabla_{\sigma}\bigg)f_{B}\bigg],
\end{equation}
where the nonmetricity scalar and boundary term in the FLRW background take the forms:
\begin{equation}\label{15}
Q=-6H^{2}, \hspace{0.4cm} B=6(3H^{2}+\dot{H}), \hspace{0.4cm} \overset{\circ}R=6(2H^{2}+\dot{H}),
\end{equation}
This geometric contribution behaves like a dynamical dark energy source, effectively modifying the evolution of the universe through its coupling to the scalar functions $Q$ and $B$. To simplify the formulation of the theory, we adopt the coincident gauge, in which the affine connection $\Gamma^{\lambda}_{\mu\nu}$ vanishes identically. This choice eliminates inertial contributions and allows all covariant derivatives to reduce to partial derivatives. Under this assumption, and using the FLRW background, one can derive the modified cosmological equations corresponding to the $f(Q, B)$ gravity framework. The generalized Friedmann equations governing the cosmic dynamics take the form:
\begin{equation}\label{16}
\kappa\rho=\frac{f}{2}+6H^{2}f_{Q}-(9H^{2}+3\dot{H})f_{B}+3H\dot{f}_{B},
\end{equation}
\begin{equation}\label{17}
\kappa p=-\frac{f}{2}-(6H^{2}+2\dot{H})f_{Q}-2H\dot{f}_{Q}+(9H^{2}+3\dot{H})f_{B}-\ddot{f}_{B}.
\end{equation}
Here, $\rho$ and $p$ denote the total energy density and pressure of the cosmic fluid, respectively. The terms involving $f_{Q}$ and $f_{B}$, as well as their derivatives, represent modifications due to the geometry of nonmetricity and its associated boundary contributions. These equations clearly reveal how the gravitational sector in $f(Q, B)$ cosmology leads to effective terms that can mimic or drive late-time acceleration, analogous to dark energy.

To proceed with a concrete cosmological analysis, we adopt a specific functional form of the gravitational action in the symmetric teleparallel framework, given by 
\begin{equation}\label{18}
f(Q,B)=\delta Q^{2}+\beta B,
\end{equation}
where $\delta$ and $\beta$ are constants that determine the coupling strengths of the quadratic nonmetricity and linear boundary contributions, respectively. This form is chosen as a natural extension of the symmetric teleparallel equivalent of General Relativity (STGR), which corresponds to the linear case $f(Q)=Q$ and allows for deviations that may account for cosmic acceleration and early-Universe inflationary behavior.

The inclusion of the $Q^{2}$ term introduces higher-order contributions to the gravitational Lagrangian and serves an analogous role to the $R^{2}$ correction in the well-known Starobinsky inflation model $f(R)=R+\gamma R^{2}$ \cite{Star80}. Such quadratic extensions are known to produce rich phenomenology in both early- and late-time cosmology by modifying the effective gravitational dynamics while preserving the second-order nature of the field equations. This behavior is particularly important in high-curvature regimes, where nonlinearities in the gravitational sector are expected to dominate.

The choice of a linear term in $B$, on the other hand, plays a more subtle yet equally significant role. Since $B\equiv R+Q$, the boundary term bridges the curvature-based and nonmetricity-based descriptions of gravity. By including it linearly, we preserve the structure of the theory while allowing for an interpolation between symmetric teleparallel gravity and the standard curvature formulation. Importantly, the linearity of $B$ ensures that the field equations remain of second order in the metric tensor, avoiding the complications associated with higher-derivative terms. In this context, it is also worth mentioning that more general models have been studied where the nonmetricity scalar appears with arbitrary powers, such as the form $f(Q,C)=\alpha Q^{n}+\beta C$, with $C$ represents a boundary-like term in terms of the Hubble parameter and its time derivative \cite{ASC25,Amits24}. These models introduce additional flexibility and are useful in exploring various cosmological regimes, including phantom-like evolution, unified dark energy behavior and early inflation. Such models have recently gained attention for their ability to interpolate between different gravitational behaviors in a single framework and have been used effectively to analyze inflationary observables and late-time acceleration in the context of symmetric teleparallel gravity.
\section{Modified chaplygin gas model in $f(Q,B)$ gravity}\label{sec3}
\hspace{0.5cm} The concept of a Chaplygin-type fluid emerged from early studies in aerodynamics, first introduced by S. Chaplygin in $1904$ \cite{Chap04}. Its application to cosmology was recognized much later, particularly due to its unusual equation of state that allows it to interpolate between pressureless matter and a cosmological constant. The original Chaplygin Gas (CG) model is defined by the equation of state (EoS):
\begin{equation}\label{19}
 p=-\frac{B}{\rho},
\end{equation}
where $B$ is a positive constant. A distinctive feature of this model is that at high energy densities (typically corresponding to early times or small scale factors), it behaves like pressureless dust, $p=0$, while at low energy densities (late times or large scale factors), it asymptotically approaches the behavior of a cosmological constant, $p=-B$. This dual behavior makes it a candidate for a unified description of dark matter and dark energy within a single fluid framework.

However, despite its simplicity and conceptual appeal, the original CG model faced challenges when confronted with observational data, particularly with structure formation and the cosmic microwave background (CMB) spectrum. To enhance the model’s flexibility and adaptability, a generalization was proposed. The Generalized Chaplygin Gas (GCG) model modifies the original equation of state to the form \cite{Bento2002}:
\begin{equation}\label{20}
p=-\frac{B}{\rho^{\alpha}},
\end{equation}
where $\alpha$ is a dimensionless parameter constrained to $0<\alpha\leq1$. This generalization allows the fluid to transition more smoothly between the matter-like and dark-energy-like regimes. As $\alpha\rightarrow1$, the model approaches the original CG case, while smaller values of $\alpha$ soften the transition and can better match observational data.

Although the GCG model offered improved phenomenology, further refinements were motivated by the need to accommodate radiation-dominated behavior at early times and more accurate representations of the Universe's expansion history. In this pursuit, the Modified Chaplygin Gas (MCG) model was proposed. It incorporates a linear barotropic term into the equation of state \cite{Jamil11,Debnath11}:
\begin{equation}\label{21}
p=A\rho-\frac{B}{\rho^{\alpha}},
\end{equation}
where $A$, $B$ and $\alpha$ are positive constants, with $0\leq\alpha\leq1$. This equation allows the MCG to model a wider range of cosmic epochs. When the energy density is high (e.g., during the early universe), the term $A\rho$ dominates, and the fluid behaves as a barotropic fluid. Depending on the choice of $A$, this can emulate radiation $(A=\frac{1}{3})$, dust $(A=0)$ or stiff-matter $(A=1)$. In the low-density regime, the second term dominates, and the negative pressure leads to accelerated expansion similar to a cosmological constant.

By substituting the modified Chaplygin gas equation of state into the standard energy conservation equation for a perfect fluid in an expanding Universe,
\begin{equation}\label{22}
\dot{\rho}+3H(\rho+p)=0,
\end{equation}
One obtains a first-order nonlinear differential equation governing the evolution of the energy density with cosmic time or redshift. Rewriting the time derivative in terms of the redshift $z$ and using the functional form of $p(\rho)$, this equation can be integrated analytically. The general solution yields the following redshift-dependent form for the energy density:
\begin{equation}\label{23}
\rho(z)=\left[ \frac{B}{1+A} + C(1+z)^{3(1+A)(1+\alpha)} \right]^{\frac{1}{1+\alpha}},
\end{equation}
where $C$ is an integration constant. This expression clearly shows that the fluid interpolates between different behaviors depending on redshift.

To simplify and normalize the solution, it is convenient to define a parameter $A_{s}$, which encapsulates the ratio between the pressure-driving term $B$ and the present-day energy density of the MCG, $\rho_{0}$ as: $A_{s}=\frac{B}{(1+A)\rho^{1+\alpha}}$. Introducing this into equation (\ref{23}) allows one to recast the energy density in terms of observable quantities:
\begin{equation}\label{24}
\rho(z)=\rho_{0}\left[A_{s}+(1-A_{s})(1+z)^{3(1+A)(1+\alpha)}\right]^{\frac{1}{1+\alpha}}.
\end{equation}
By substituting the energy density expression given in equation (\ref{24}) into the modified Friedmann equation (\ref{16}) and using the specific form of the gravitational model $f(Q,B)$ which is introduced in equation (\ref{18}), we obtain the Hubble parameter as a function of redshift:
\begin{equation}\label{25}
H(z)=H_{0}\left[A_{s}+(1-A_{s})(1+z)^{3(1+A)(1+\alpha)}\right]^{\frac{1}{4(1+\alpha)}}.
\end{equation}
Here, $H_{0}$ represents the present value of the Hubble parameter and is related to the present-day energy density $\rho_{0}$ of the Modified Chaplygin Gas through the relation: $H_{0}^{4}=-\frac{\rho_{0}}{54\delta}$. It is obtained that $\delta<0$. To constrain the free parameters of the model—namely $H_{0}$, $A_{s}$, $A$ and $\alpha$—we utilize current observational datasets, including Hubble parameter measurements, Supernovae Type Ia data, and Baryon Acoustic Oscillation (BAO) observations. By performing a statistical analysis using these datasets, we aim to obtain the best-fit values of the model parameters and evaluate the viability of the Modified Chaplygin Gas in the context of $f(Q,B)$ gravity.
\section{Observational constraints and statistical analysis}\label{sec4}
\hspace{0.5cm} In this section, we aim to place constraints on the parameters of our cosmological model using the most recent observational data. For this purpose, we employ a Bayesian inference framework utilizing the Markov Chain Monte Carlo (MCMC) technique, implemented via the open-source Python library emcee \cite{Mackey13}. This method allows for an effective examination of the parameter space and facilitates the determination of credible intervals for the parameters under investigation. The likelihood function employed in this analysis is constructed from a standard Gaussian form, where the total probability is proportional to:
\begin{equation}\label{26}
\mathcal{L}\propto e^{-\frac{\chi^{2}}{2}},
\end{equation}
with $\chi^{2}$ denoting the chi-square function associated with the respective dataset. The parameter priors used in this study are: $50<H_{0}<100$, $0<A_{s}<3$, $0<A<1$ and $0<\alpha<1$. To ensure a comprehensive sampling of the parameter space, we run the MCMC analysis using $100$ walkers and $1500$ iterations per chain. Each walker is initialized randomly within the chosen priors, and convergence is checked by monitoring the autocorrelation time of the chains. The total chi-square function is constructed by summing the contributions from each independent dataset. In what follows, we describe the construction of the chi-square function for each of the datasets used in this analysis: the Hubble parameter measurements, Baryon Acoustic Oscillations (BAO), the DESI DR2 BAO sample and the Pantheon+ Type Ia Supernovae compilation.
\subsection{Hubble dataset}\label{sec4.1}
\hspace{0.5cm} For the Hubble dataset, we use a collection of $46$ independent measurements of the Hubble parameter $H(z)$ which is obtained through the cosmic chronometer (CC) approach. This technique is based on measuring the differential age of passively evolving early-type galaxies across various redshifts, offering a model-independent method to determine $H(z)=-(1+z)^{-1}\frac{dz}{dt}$. which is derived from the FLRW metric. These measurements cover a wide range of redshifts and are compiled from numerous observational surveys and studies utilizing high-precision spectroscopic data. To evaluate the consistency of the model with the observational data, we define the chi-square statistic for the Hubble dataset as:
\begin{equation}\label{27}
\chi^{2}_{H(z)}=\Delta H^{T}C^{-1}_{H(z)}\Delta H,
\end{equation}
where $\Delta H=H_{model}(z)-H_{obs}(z)$ is the residual vector between the theoretical and observed Hubble values and $C_{H(z)}$ is the covariance matrix representing uncertainties and correlations in the dataset. This formulation allows us to statistically assess the likelihood of the model predictions given the observed data, serving as a foundation for the full MCMC analysis.
\subsection{Baryon acoustic oscillation data}\label{sec4.2}
\hspace{0.5cm} Baryon Acoustic Oscillations (BAO) provides a robust standard ruler in cosmology, rooted in the sound waves that propagated through the primordial plasma before recombination. The characteristic scale imprinted by these acoustic oscillations is related to the comoving sound horizon at the drag epoch, denoted by $r_{d}$, which is defined as:
\begin{equation}\label{28}
r_{d}= \int_{z_d}^\infty \frac{c_s(z)}{H(z)} dz,
\end{equation}

where $c_s(z)$ is the sound speed in the photon-baryon fluid, and $z_d$ corresponds to the redshift of the drag epoch (typically around 1060). While $r_d$ is fixed in standard $\Lambda$CDM, we treat it as a free parameter in our analysis to account for possible deviations arising from modified gravity.

In our work, we utilize a total of $15$ BAO measurements \cite{DESI24,AG25,AGA25} and the most recent DESI Data Release $2$ (DR2) measurements \cite{Karim2025}. The DESI DR2 dataset provides high-precision constraints on large-scale structure across various tracers, such as the Bright Galaxy Sample (BGS), Luminous Red Galaxies (LRGs), Emission Line Galaxies (ELGs) and high-redshift quasars, as well as Lyman-$\alpha$ forest absorption systems.

These observations are reported in terms of various cosmological distance ratios, including:\\
The Hubble distance: $D_H(z) = \frac{c}{H(z)}$,\\
The comoving angular diameter distance: $D_M(z) = \int_0^z \frac{c}{H(z')} dz'$,\\
The volume-averaged effective distance: $D_V(z) = \left[ z D_M^2(z) D_H(z) \right]^{1/3}$.

The model predictions are tested against the observed ratios $\frac{D_M}{r_d}$, $\frac{D_H}{r_d}$, $\frac{D_V}{r_d}$ and $\frac{D_M}{D_H}$, as reported by the respective datasets. The corresponding chi-squared function for BAO is defined as:
\begin{equation}\label{29}
\chi^2_{BAO} = \Delta D^T C^{-1} \Delta D,
\end{equation}
where $\Delta D=D_{obs}-D_{th}$, and $C^{-1}$ is the inverse of the covariance matrix provided with the data.
\subsection{Pantheon+ dataset}\label{sec4.3}
\hspace{0.5cm} Type Ia supernovae (SNe Ia) have been extensively used as dependable tools in cosmology because of their consistent peak brightness, making them ideal standard candles for determining distances to faraway galaxies. These explosive events provide a direct way to trace the universe's expansion history, particularly when paired with accurate redshift measurements. One of the most detailed and expansive collections of SNe Ia data is the Pantheon+ compilation, which includes a total of $1701$ supernovae spanning the redshift range $z \in [0.001, 2.3]$. This rich dataset greatly enhances the statistical robustness of late-time cosmological studies and allows for a more detailed examination of dark energy and the accelerating expansion of the Universe \cite{Kowa08,RA10,Suzu12,Beto14,Scolnic18}. 

The distance modulus $\mu(z)$, which links the observed apparent magnitudes to theoretical distance estimates, is given by the formula:
\begin{equation}\label{30}
\mu(z) = 5 \log_{10} \left( \frac{d_L(z)}{1\, Mpc} \right) + 25,
\end{equation}
where $d_L(z)$ represents the luminosity distance. In a spatially flat FLRW cosmological model, this quantity is calculated using the following expression:
\begin{equation}\label{31}
d_L(z) = c (1 + z) \int_0^z \frac{dz'}{H(z'; H_{0},A_{s},A,\alpha)},
\end{equation}
To incorporate both statistical and systematic uncertainties in the supernova data, we employ the complete covariance matrix $C_{Pantheon+}$. The chi-squared statistic used to evaluate the goodness-of-fit of the model is expressed as:
\begin{equation}\label{32}
\chi^2_{Pantheon+} = \Delta\mu^\mathrm{T} C_{Pantheon+}^{-1} \Delta\mu,
\end{equation}
where the residual vector $\Delta\mu$ captures the difference between the observed and predicted distance moduli, defined as:
\begin{equation}\label{33}
\Delta\mu_j = m_{B_j} - M - \mu_{model}(z_j),
\end{equation}
Here, $m_{B_j}$ denotes the observed apparent magnitude of the $j$-th supernova, $M$ represents the absolute magnitude, and $\mu_{model}(z_j)$ is the theoretically computed distance modulus for the given redshift according to the chosen cosmological model.

An important enhancement in the Pantheon+ dataset is the incorporation of Cepheid-calibrated host galaxies, which helps resolve the degeneracy between the absolute magnitude $M$ and the Hubble constant $H_0$. For supernovae occurring in these Cepheid-hosting galaxies, the observed distance modulus is substituted with independently determined values $\mu_j^{\text{Ceph}}$. As a result, the residual vector is modified as follows:
\begin{equation}\label{34}
\Delta\mu'_j = 
\begin{cases}
m_{B_j} - M - \mu_j^{Ceph}, & \text{for Cepheid-hosted SNe}, \\
m_{B_j} - M - \mu_{model}(z_j), & \text{otherwise}.
\end{cases}
\end{equation}
This adjustment enables a simultaneous and consistent determination of both $M$ and $H_0$, thereby enhancing the accuracy and reliability of cosmological parameter estimates based on the Pantheon+ supernova sample.
\subsection{Combined likelihood analysis and parameter estimation}\label{sec4.4}
\hspace{0.5cm} In order to place stringent constraints on the parameters of our cosmological model, we perform a joint analysis by combining multiple observational datasets: the Hubble parameter measurements, Baryon Acoustic Oscillation (BAO) data (including DESI DR2) and the Pantheon+ compilation of Type Ia Supernovae. Each dataset provides independent and complementary information about the expansion history of the Universe, and their combination allows for a more comprehensive and reliable estimation of the model parameters.

The total chi-square function used in our statistical analysis is the sum of the individual contributions from each dataset:
\begin{equation}\label{35}
\chi^{2}_{total}=\chi^{2}_{H(z)}+\chi^{2}_{BAO}+\chi^{2}_{SNe Ia},
\end{equation}
This total likelihood is used within a Bayesian framework, and parameter constraints are derived via Markov Chain Monte Carlo (MCMC) sampling. The analysis explores a four-dimensional parameter space involving the Hubble constant $H_0$, the modified Chaplygin gas parameters $A_s$, $A$, and the exponent $\alpha$.

\begin{figure}[hbt!]
  \centering
  \includegraphics[scale=0.33]{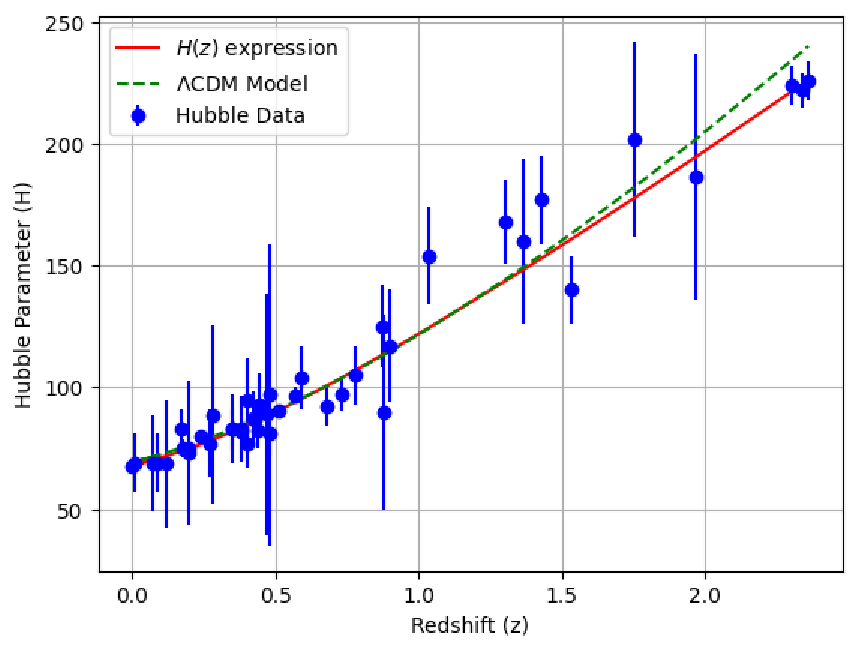}~
  \includegraphics[scale=0.33]{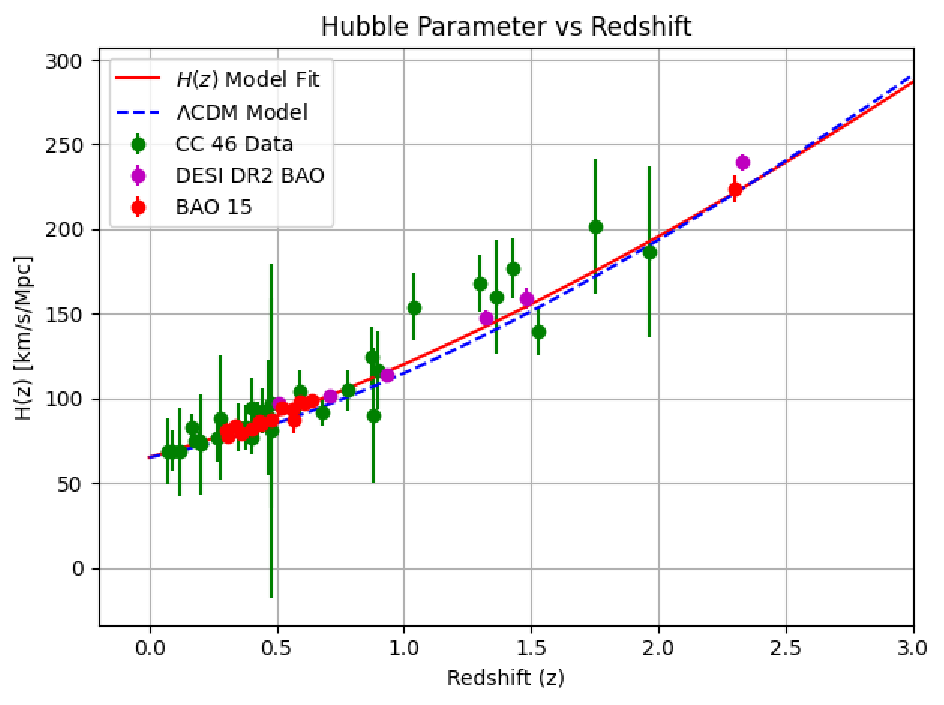}~~
  \includegraphics[scale=0.28]{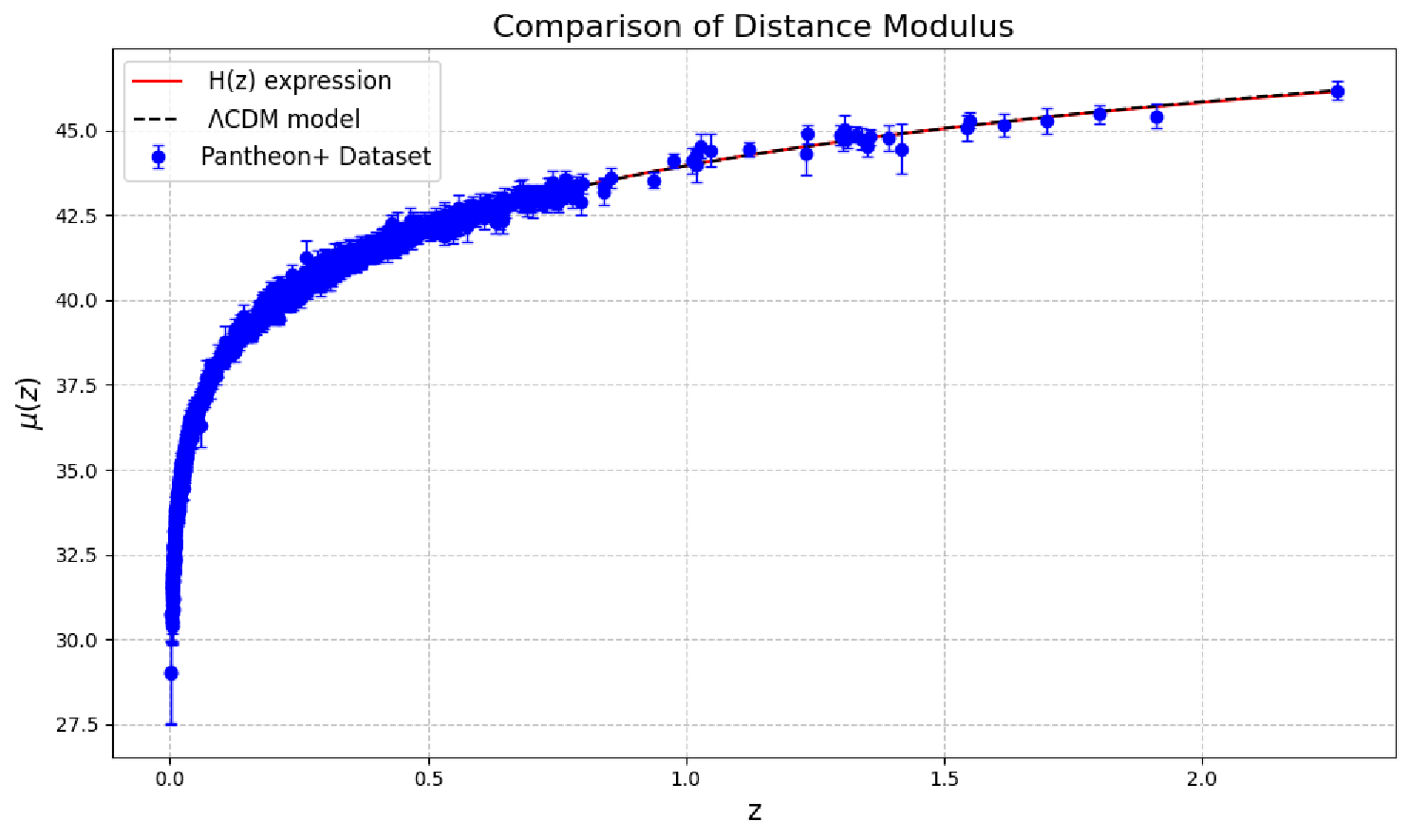}
  \caption{Error bar analysis depicting variations in parameter estimates from dataset combinations.}\label{fig:f1}
\end{figure}

\begin{figure}[hbt!]
  \centering
  \includegraphics[scale=0.38]{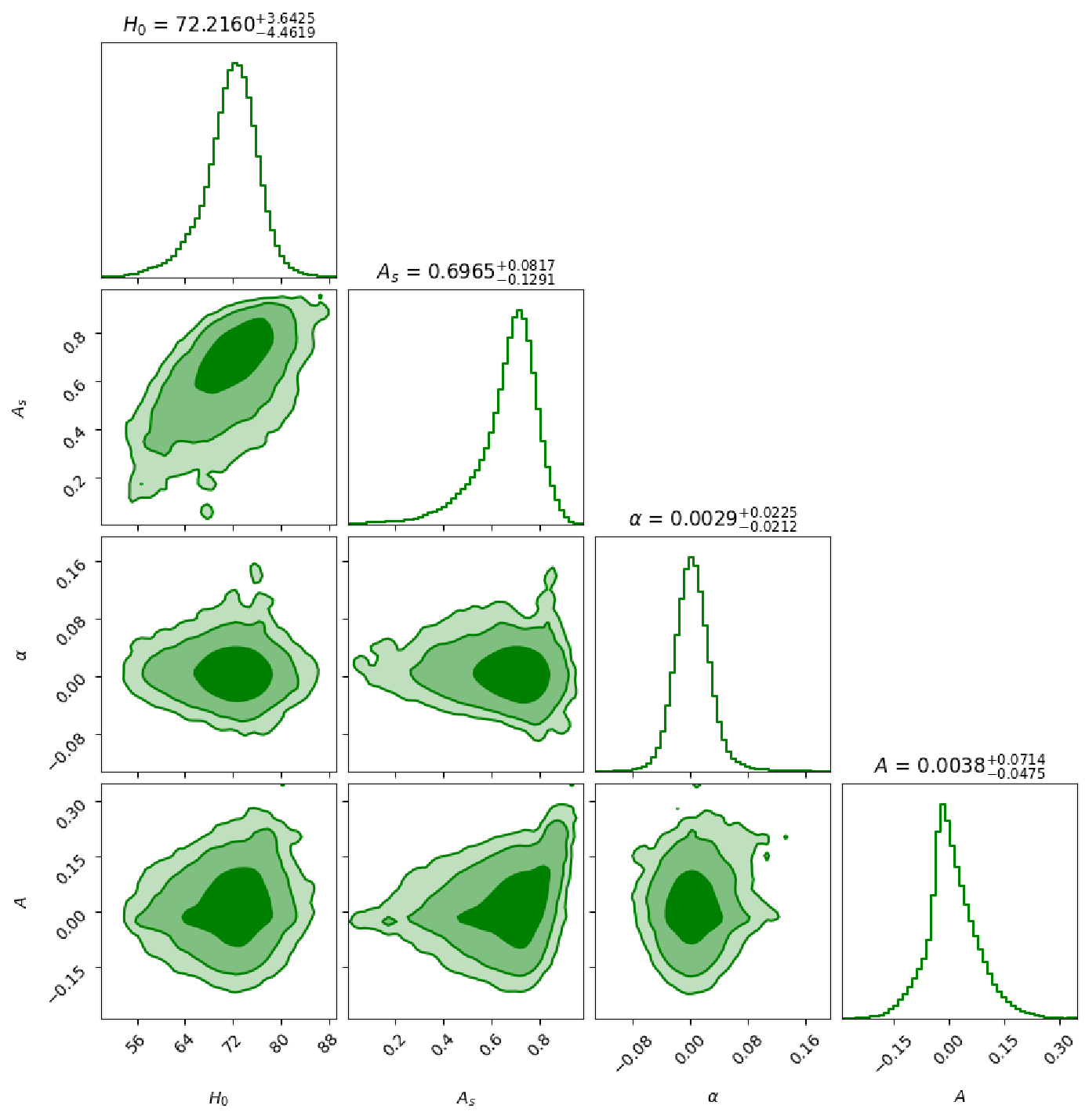}
  \caption{Joint confidence contours for $(H_{0},A_{s},A,\alpha)$ showing $1\sigma$, $2\sigma$ and $3\sigma$ regions from combined datasets.}\label{fig:f2}
\end{figure}

We present the posterior distributions of the parameters through one-dimensional marginalized probability curves, along with two-dimensional contour plots for relevant parameter pairs such as $(H_0, A_s)$, $(H_0, \alpha)$, $(A_s, A)$ and $(A, \alpha)$. The contour plots show confidence regions corresponding to $1\sigma$ $(68.27\%)$, $2\sigma$ $(95.45\%)$ and $3\sigma$ $(99.73\%)$ levels. These results, illustrated in Figures \ref{fig:f1} and \ref{fig:f2}, highlight the effectiveness of multi-probe constraints in reducing parameter degeneracies and improving the precision of cosmological model testing.

From the MCMC analysis using the joint observational dataset (Hubble, BAO, and Pantheon+), the best-fit values for the model parameters are obtained as follows: $H_0=72.2160^{+3.6425}_{-4.4619}$ km/s/Mpc, $\quad A_s = 0.6965^{+0.0817}_{-0.1291}, \quad \alpha = 0.0029^{+0.0225}_{-0.0212}, \quad A = 0.0038^{+0.0714}_{-0.0475}$. From the expression $H_{0}^{4}=-\frac{\rho_{0}}{54\delta}$, by using above observational constrained present values  of $H_{0}$ and $\rho_{0}=1$, we obtain $\delta=-6.8\times10^{-11}$.

The Hubble constant $H_0$, as previously mentioned, aligns well with the SH0ES local determination ($H_0 \approx 73.3 \pm 1.0\,\mathrm{km\,s^{-1}\,Mpc^{-1}}$) \cite{Riess2022}, while remaining within $2\sigma$ tension with the Planck 2018 CMB-based inference under the $\Lambda$CDM model ($H_0 \approx 67.4 \pm 0.5 \,\mathrm{km\,s^{-1}\,Mpc^{-1}}$) \cite{Planck2018}. This agreement suggests that our modified gravity model has the potential to alleviate the current Hubble tension by allowing more flexibility in the late-time expansion behavior.

The parameter $A_s$, which governs the relative contribution of the MCG component at late times, is found to be approximately 0.70. This is close to the expected dark energy fraction today in standard cosmology ($\Omega_\Lambda \sim 0.7$), implying that the geometrically-induced effective energy density in $f(Q,B)$ gravity can play a role analogous to dark energy. The result supports the viability of the MCG model in recovering the observed accelerated expansion of the universe without requiring an explicit cosmological constant.

The Chaplygin index $\alpha$ is notably close to zero with small uncertainty, indicating that the model behaves very similarly to the $\Lambda$CDM model. In particular, $\alpha \to 0$ reduces the MCG equation of state to the standard Generalized Chaplygin Gas (GCG), and further to a $\Lambda$CDM-like scenario when $A \to 0$. The best-fit value $\alpha = 0.0029$ therefore suggests only a mild deviation from a cosmological constant-like behavior, consistent with observational constraints on time-varying dark energy.

The barotropic coefficient $A$, controlling the early-time behavior of the fluid, is also found to be small and positive. This suggests that the MCG component behaves nearly as pressureless matter during the matter-dominated era, allowing a smooth transition from decelerated to accelerated expansion. The small deviation from $A = 0$ allows for subtle improvements in fitting the expansion history, while remaining consistent with structure formation data.

Overall, the obtained values of $A_s$, $\alpha$, and $A$ indicate that the MCG in $f(Q,B)$ gravity successfully interpolates between a dust-like fluid at high redshift and a dark energy-like component at low redshift. This provides a natural mechanism for unifying the cosmic phases of deceleration and acceleration, without the need for a fine-tuned cosmological constant. The results show that the model remains observationally viable and offers a promising alternative to standard $\Lambda$CDM cosmology within the framework of symmetric teleparallel modified gravity.
\section{Evolution of cosmological parameters in $f(Q,B)$ gravity}\label{sec5}
\hspace{0.5cm} In this section, we analyze the evolution of essential cosmological parameters derived from the modified Chaplygin gas model within the context of $f(Q, B)$ gravity. These parameters play a vital role in understanding the Universe's expansion history and assessing the model's effectiveness in capturing both the early and late stages of cosmic evolution.
\subsection{Deceleration parameter}\label{sec5.1}
\hspace{0.5cm} The deceleration parameter $q(z)$ plays a vital role in characterizing the expansion dynamics of the Universe. It is defined as $q(z)=-1-\frac{\dot{H}}{H^{2}}$, where a positive value of $q$ indicates decelerated expansion, typical of the matter-dominated era, while negative values correspond to accelerated expansion as expected in the current epoch. In the standard $\Lambda$CDM cosmology, the Universe undergoes a transition from a decelerating to an accelerating phase at a redshift $z_{tr}\sim0.6-0.8$, which is consistent with Type Ia supernovae and BAO observations. In the context of our $f(Q,B)$ gravity model with Modified Chaplygin Gas, we derive an explicit form for $q(z)$ using the expression for the Hubble parameter obtained in equation (\ref{25}).
\begin{equation}\label{36}
q(z)=-1+\frac{3(1-A_{s})(1+A)(1+z)^{3(1+\alpha)(1+A)}}{4\left[A_{s}+(1-A_{s})(1+z)^{3(1+\alpha)(1+A)}\right]^{\frac{1}{4(1+\alpha)}}}.
\end{equation} 
\begin{figure}[hbt!]
  \centering
  \includegraphics[scale=0.4]{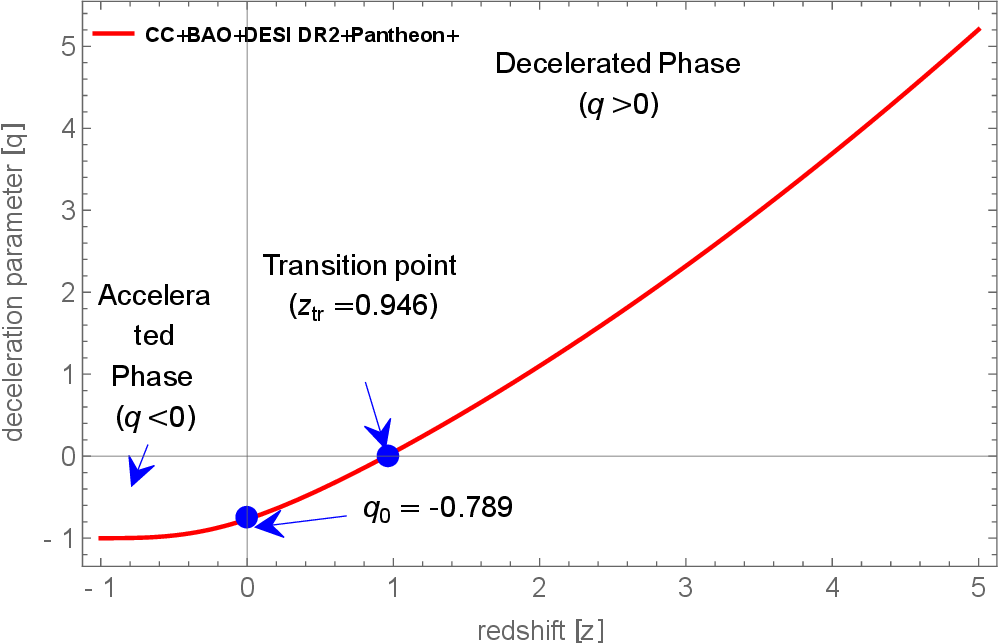}
  \caption{Plot of the deceleration parameter $q(z)$ versus redshift in the $f(Q,B)$ gravity model with Modified Chaplygin Gas.}\label{fig:f3}
\end{figure}

The evolution of $q(z)$ is shown in Figure \ref{fig:f3}, where we explore its behavior across different redshifts. The plot reveals that the Universe starts in a decelerated phase, with the deceleration parameter taking a positive value at high redshifts (early times). This phase corresponds to a matter-dominated regime necessary for the formation of large-scale structure. As the Universe evolves, $q(z)$ gradually decreases, indicating a weakening of deceleration due to the increasing influence of the Chaplygin-like fluid. A key feature of this model is the smooth transition from deceleration to acceleration at the redshift $z_{tr}\approx0.946$, which is in good agreement with recent observational constraints on the transition redshift. Beyond this point, $q(z)$ continues to decline and asymptotically approaches $-1$, which indicates that the Universe evolves toward a de Sitter-like phase characterized by exponential expansion. At present epoch $z=0$, the model predicts $q_{0}=-0.789$ which is consistent with observational bounds from cosmic chronometers and supernovae datasets. This value confirms that the current Universe is undergoing a period of accelerated expansion, closely resembling the late-time behavior of the $\Lambda$CDM model but emerging naturally from the geometry-induced dynamics of $f(Q,B)$ gravity \cite{SSS2025,AS25}.

This continuous and observationally consistent evolution of the deceleration parameter demonstrates the capability of the MCG model in $f(Q,B)$ gravity to account for the entire cosmic timeline — from matter domination to present-day acceleration — without requiring a fine-tuned cosmological constant.
\subsection{Evolution of energy density and pressure}\label{sec5.2}
\hspace{0.5cm} In this section, we analyze the behavior of the energy density $\rho(z)$ and pressure $p(z)$ for the MCG in the framework of $f(Q,B)$ gravity. These two quantities govern the dynamics of the cosmic fluid and play a central role in determining the expansion history of the Universe. Using the derived expressions for $\rho(z)$ from the energy conservation equation and substituting into the modified field equations of $f(Q,B)$ gravity, we obtain explicit forms of $\rho(z)$ and $p(z)$ that reflect both geometrical and thermodynamical effects of the MCG fluid.
\begin{equation}\label{37}
\rho(z)=-54\delta H_{0}^{4}\left[A_{s}+(1-A_{s})(1+z)^{3(1+A)(1+\alpha)}\right]^{\frac{1}{(1+\alpha)}},
\end{equation}
\begin{equation}\label{38}
p(z)=18\delta H_{0}^{4}\left[A_{s}+(1-A_{s})(1+z)^{3(1+A)(1+\alpha)}\right]^{\frac{1}{(1+\alpha)}}\bigg[3-\frac{(1-A_{s})(1+A)(1+z)^{3(1+A)(1+\alpha)}}
{\left[A_{s}+(1-A_{s})(1+z)^{3(1+A)(1+\alpha)}\right]^{\frac{1}{4(1+\alpha)}}}\bigg].
\end{equation}
\begin{figure}[hbt!]
  \centering
  \includegraphics[scale=0.4]{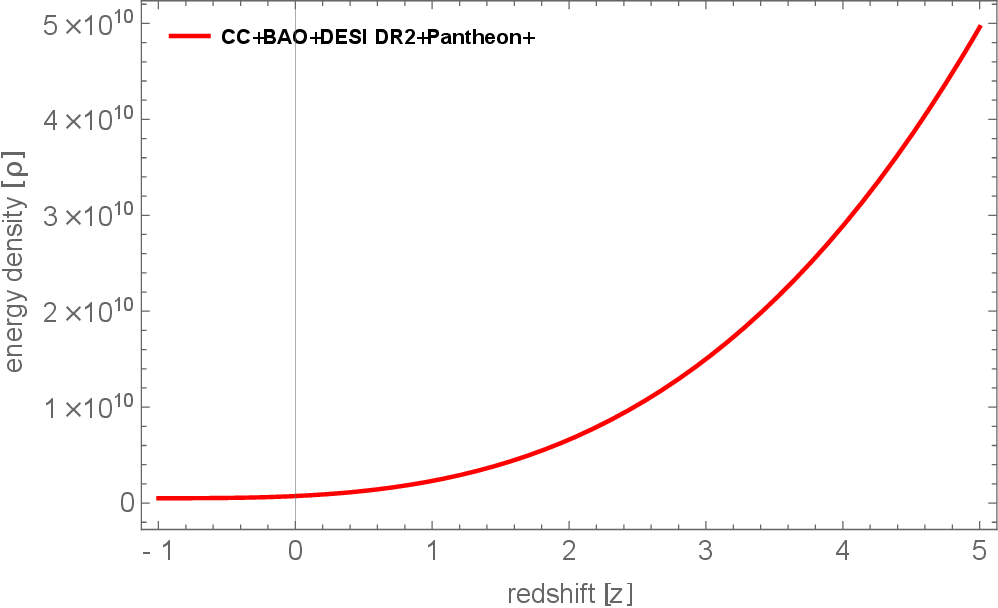}~~~
  \includegraphics[scale=0.4]{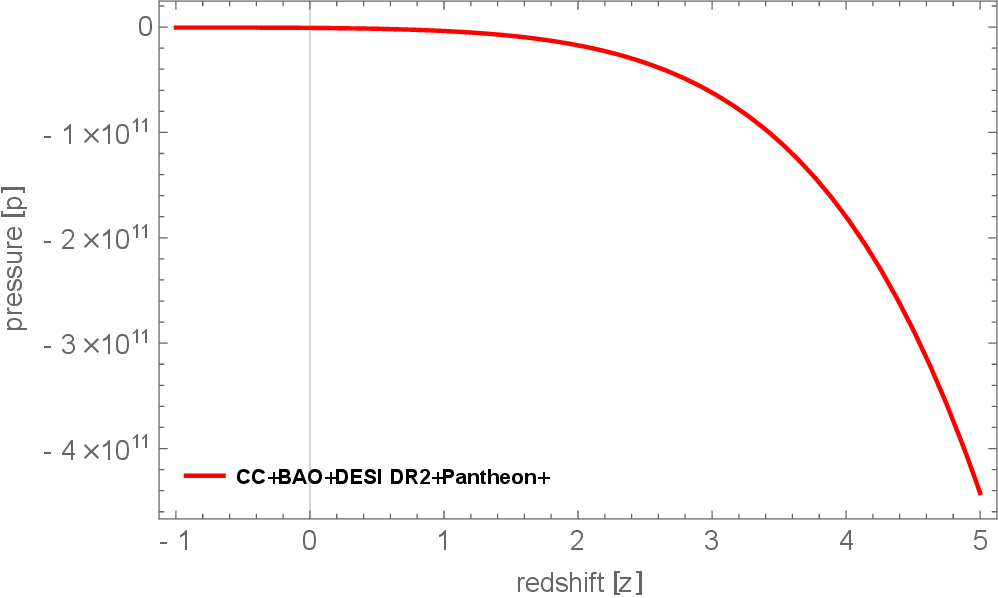}
  \caption{Evolution of energy density and pressure as functions of redshift for $\delta=-0.5$.}\label{fig:f4}
\end{figure}

Figure \ref{fig:f4} illustrates the evolution of these quantities across cosmic time. The plot clearly shows that the energy density remains positive throughout the redshift range, confirming the physical viability of the model and its consistency with the weak energy condition. In contrast, the pressure is negative during the recent and late-time cosmic epochs, a hallmark of dark energy–like behavior. This negative pressure is the driving force behind the current acceleration of the Universe and emerges naturally from the nonlinear structure of the MCG in the $f(Q,B)$ background.
\subsection{Evolution of the EoS parameter}\label{sec5.3}
\hspace{0.5cm} The equation of state (EoS) parameter $\omega(z)$, defined as the ratio of pressure to energy density, $\omega(z) = \frac{p(z)}{\rho(z)}$ which serves as a fundamental diagnostic tool to characterize the nature of cosmic fluids and their influence on the expansion dynamics of the Universe. It distinguishes between different cosmological regimes such as radiation ($\omega = \frac{1}{3}$), matter ($\omega = 0$), quintessence-like dark energy ($-1 < \omega < -0.33$) and the cosmological constant ($\omega = -1$). Using the derived expressions for the energy density $\rho(z)$ and pressure $p(z)$ in our model, we compute the redshift-dependent evolution of the EoS parameter $\omega(z)$ for the Modified Chaplygin Gas in the context of $f(Q, B)$ gravity.
\begin{equation}\label{39}
\omega(z)=-1+\frac{(1-A_{s})(1+A)(1+z)^{3(1+\alpha)(1+A)}}{3\left[A_{s}+(1-A_{s})(1+z)^{3(1+A)(1+\alpha)}\right]^{\frac{1}{4(1+\alpha)}}}.
\end{equation}
\begin{figure}[h!]
  \centering
  \includegraphics[scale=0.4]{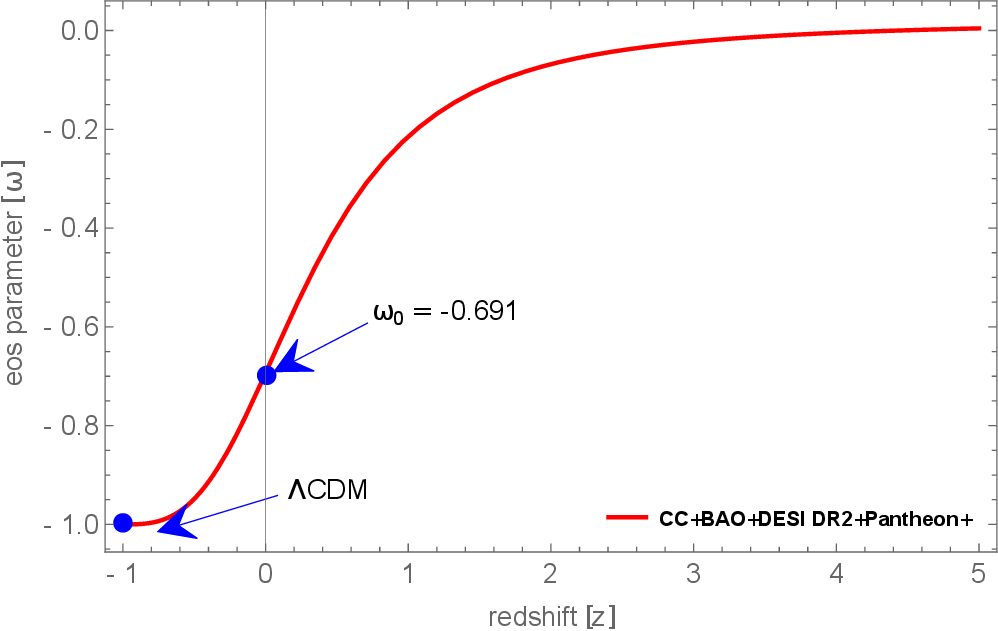}
  \caption{Evolutionary trend of $\omega(z)$ in the framework of $f(Q, B)$ gravity with MCG for $\delta=-0.5$.}\label{fig:f5}
\end{figure}

Figure \ref{fig:f5} displays the variation of $\omega(z)$ as a function of redshift. The plot exhibits a rich structure consistent with the expected cosmic timeline. At high redshift values ($z\gg 1$), the EoS parameter begins at a small positive value $\omega(z\gg1) \approx 0.0045$, which indicates a matter-like behavior during the early Universe, essential for structure formation. As the Universe evolves, $\omega(z)$ gradually decreases, crossing into the negative region. This decline marks the onset of dark energy dominance, reflecting the transition from a decelerating to an accelerating expansion phase. In the low redshift regime ($z<1$), the EoS parameter continues to decrease and asymptotically approaches $\omega_0 \approx -0.691$, at the present epoch, which indicates a clear departure from pressureless matter toward a dark energy–dominated state. At late times ($z \to -1$), the EoS parameter tends toward $\omega=-1$, mimicking a cosmological constant and suggesting the Universe will evolve into a de Sitter-like phase of eternal acceleration. The full evolution of $\omega(z)$ hence captures all key cosmic epochs — a near matter-dominated past, a smooth transition period and a dark energy–dominated future — validating the capability of the MCG model in $f(Q, B)$ gravity to effectively unify the different phases of cosmic history without invoking separate components.
\section{Energy conditions in $f(Q,B)$ gravity with MCG}\label{sec6}
\begin{figure}[h!]
  \centering
  \includegraphics[scale=0.42]{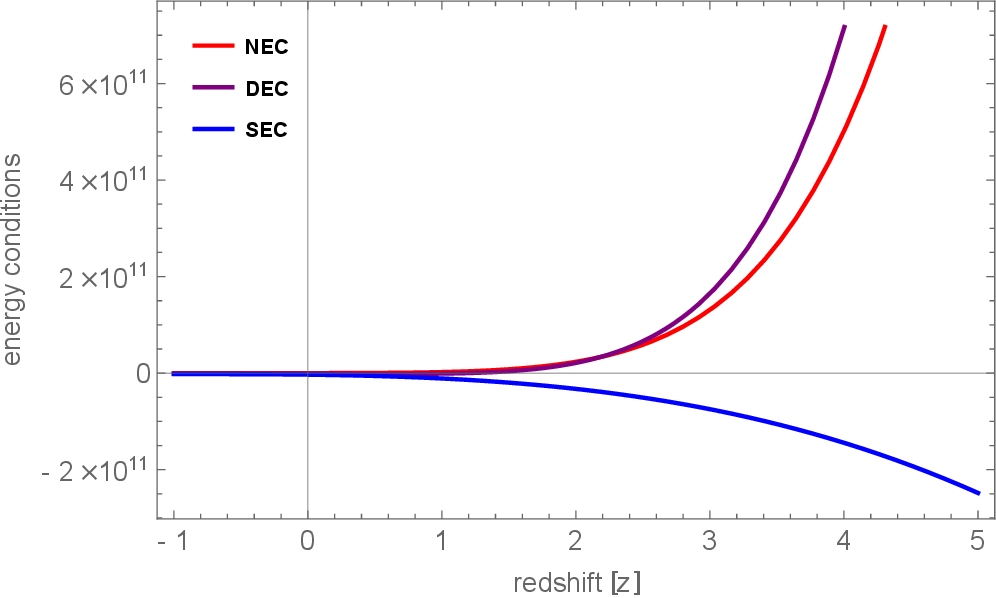}
  \caption{Redshift evolution of the energy conditions for the MCG in $f(Q, B)$ gravity.}\label{fig:f6}
\end{figure}

\hspace{0.5cm} Energy conditions serve as powerful theoretical tools in general relativity and modified gravity theories, offering insights into the physical plausibility of matter-energy content and the nature of gravitational interactions. By imposing certain inequalities on the energy-momentum tensor, these conditions allow us to test whether a cosmological model is consistent with basic physical expectations such as the attractive nature of gravity, causality and positive energy density. In the context of our $f(Q,B)$ gravity model with the MCG, we evaluate the following standard energy conditions: Null Energy Condition (NEC): $\rho+p\geq0$, Dominant Energy Condition (DEC): $\rho-p\geq0$ and Strong Energy Condition (SEC): $\rho+3p\geq0$. Using the derived expressions for $\rho(z)$ and $p(z)$, we compute these quantities as functions of redshift. The evolution of these conditions is shown in Figure \ref{fig:f6}.
\begin{equation}\label{40}
\rho+p=-18\delta H_{0}^{4}(1-A_{s})(1+A)(1+z)^{3(1+\alpha)(1+A)}\left[A_{s}+(1-A_{s})(1+z)^{3(1+A)(1+\alpha)}\right]^{\frac{3}{4(1+\alpha)}},
\end{equation}
\begin{equation}\label{41}
\rho-p=18\delta H_{0}^{4}\left[A_{s}+(1-A_{s})(1+z)^{3(1+A)(1+\alpha)}\right]^{\frac{1}{(1+\alpha)}}\bigg[6+\frac{(1-A_{s})(1+A)(1+z)^{3(1+A)(1+\alpha)}}
{\left[A_{s}+(1-A_{s})(1+z)^{3(1+A)(1+\alpha)}\right]^{\frac{1}{4(1+\alpha)}}}\bigg].
\end{equation}
\begin{equation}\label{42}
\rho+3p=54\delta H_{0}^{4}\left[A_{s}+(1-A_{s})(1+z)^{3(1+A)(1+\alpha)}\right]^{\frac{1}{(1+\alpha)}}\bigg[2-\frac{(1-A_{s})(1+A)(1+z)^{3(1+A)(1+\alpha)}}
{\left[A_{s}+(1-A_{s})(1+z)^{3(1+A)(1+\alpha)}\right]^{\frac{1}{4(1+\alpha)}}}\bigg].
\end{equation}

The plot reveals that both the NEC and DEC are satisfied throughout cosmic history. They begin with high positive values at early times $(z\gg1)$ and gradually decrease, approaching zero as redshift tends to $z\rightarrow-1$, which corresponds to the far future. This behavior reflects that the model respects fundamental energy requirements and causality during the matter-dominated and dark-energy–dominated phases. However, the SEC behaves differently. It starts with negative values at high redshift, remains negative during the entire evolution and asymptotically approaches $-1$ as $z\rightarrow-1$. This persistent violation of the SEC is expected and physically consistent with the accelerating expansion of the Universe, as confirmed by numerous cosmological observations such as Type Ia supernovae and Planck data. The violation of the SEC is often regarded as a signature of dark energy and our model reproduces this naturally through the nonlinear Chaplygin gas dynamics in $f(Q,B)$ gravity. Hence, the energy condition analysis further supports the viability of our model in describing the transition from decelerated to accelerated expansion, while maintaining physically acceptable behavior of the cosmic fluid.
\section{Dynamical analysis in the $\omega-\omega'$ plane}\label{sec7}
\hspace{0.5cm} The $\omega-\omega'$ plane was introduced by Caldwell and Linder \cite{RRC05}, is a powerful diagnostic tool to study the dynamical behavior of dark energy models. It offers a phase-space perspective by plotting the EoS parameter $\omega$ against its redshift derivative $\omega'=\frac{d\omega}{d\ln a}=-(1+z)\frac{d\omega}{dz}$. This tool helps distinguish between different classes of dark energy models—especially freezing and thawing types—based on the trajectory of evolution in the $\omega-\omega'$ space. Substituting our expression of $\omega(z)$ for the MCG in the context of $f(Q,B)$ gravity, we compute the evolution trajectory in the $\omega-\omega'$ plane.
\begin{equation}\label{43}
\omega'=\frac{(1-A_{s})(1+\alpha)(1+A)^{2}(1+z)^{3(1+\alpha)(1+A)}}{\left[A_{s}+(1-A_{s})(1+z)^{3(1+A)(1+\alpha)}\right]^{\frac{1}{4(1+\alpha)}}}-
\frac{(1-A_{s})^{2}(1+\alpha)(1+A)^{2}(1+z)^{3(1+\alpha)(1+A)}}{4\left[A_{s}+(1-A_{s})(1+z)^{3(1+A)(1+\alpha)}\right]^{\frac{1}{4(1+\alpha)}+1}}.
\end{equation}
\begin{figure}[hbt!]
  \centering
  \includegraphics[scale=0.44]{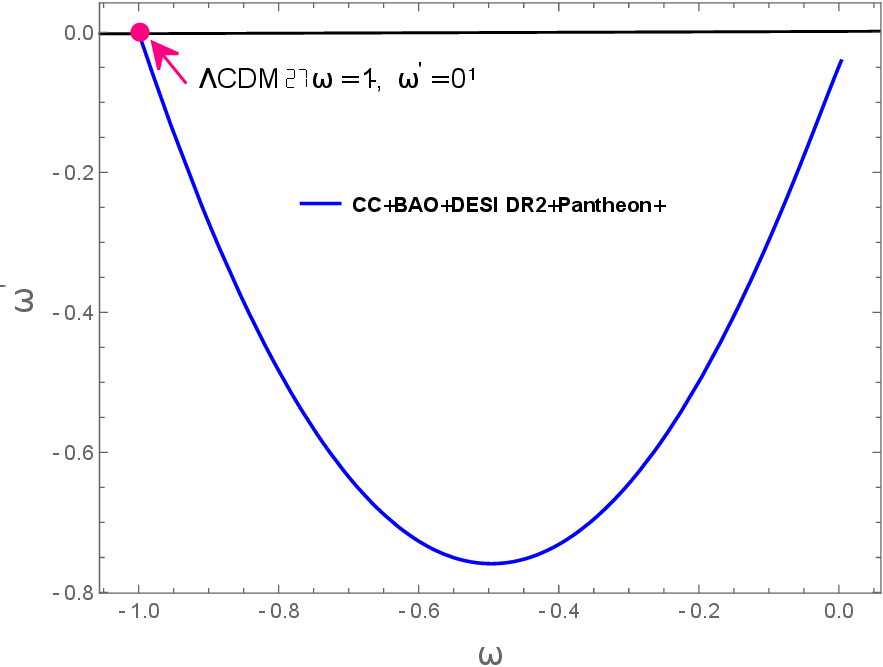}
  \caption{Trajectory in the $\omega-\omega'$ plane for the MCG in $f(Q, B)$ gravity.}\label{fig:f7}
\end{figure}

Figure \ref{fig:f7} illustrates the path of this evolution. The trajectory begins in the matter-like regime with $(\omega,\omega')\approx(0.0016,-0.044)$, which indicates nearly pressureless behavior with mild evolution. As the redshift decreases, the curve enters a dynamical transition phase, where $\omega$ becomes more negative and $\omega'$ sharply decreases, which reaches a minimum at $(\omega,\omega')\approx(-0.5107,-0.7594)$, which signifies a strong evolution of the dark energy component. After this turning point, the system enters a freezing phase, where $\omega$ continues to decline slowly and $\omega'$ increases, leading the trajectory to asymptotically approach the cosmological constant behavior at $(\omega,\omega')=(1,0)$. This convergence to the $\Lambda$CDM point demonstrates that the MCG model in $f(Q,B)$ gravity behaves like a cosmological constant in the far future, while maintaining dynamical properties in the recent past. The path of evolution also lies predominantly in the freezing region of the $\omega-\omega'$ plane, which suggests a gradual slowing down of the EoS evolution, consistent with observational expectations.
\section{Age of the Universe}\label{sec8}
\hspace{0.5cm} The age of the Universe is one of the most critical benchmarks for evaluating the viability of cosmological models. Any consistent model must predict an age that aligns with astrophysical and cosmological observations, such as those from globular clusters, the CMB and supernovae. In our modified gravity framework based on $f(Q,B)$ gravity with a MCG equation of state, the age of the Universe $t_{0}$ is computed using the redshift-dependent Hubble parameter:
\begin{equation}\label{44}
t_{0}-t=\int_{0}^{z}\frac{dz}{(1+z)H(z)},
\end{equation}
\begin{figure}[hbt!]
  \centering
  \includegraphics[scale=0.4]{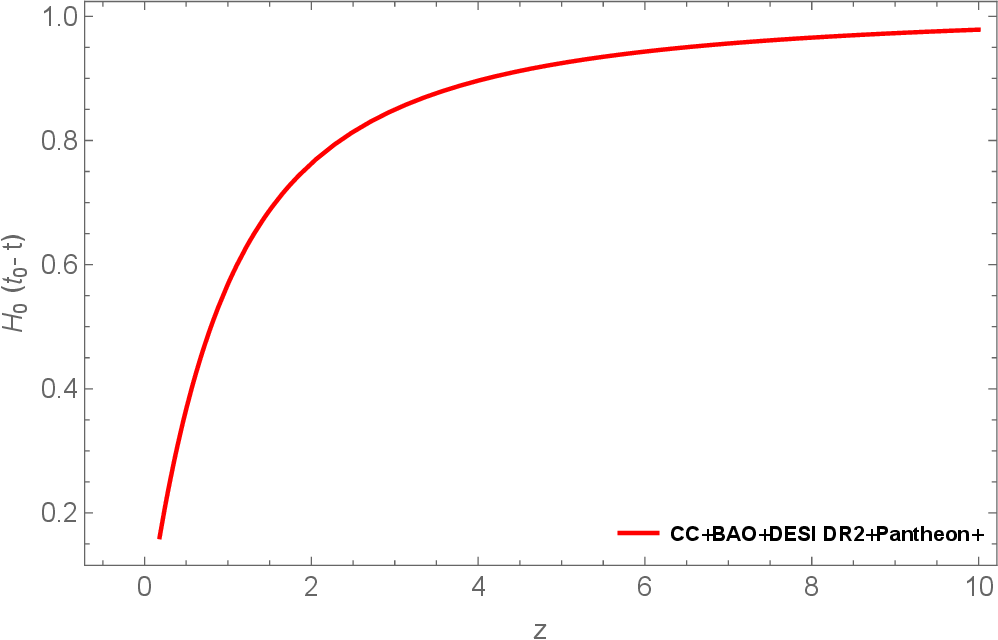}
  \caption{Evolution of the $H_0(t_0 - t)$ as a function of redshift $z$.}\label{fig:f8}
\end{figure}

Using the best-fit model parameters derived from the combined observational datasets (Pantheon+, Hubble 46, BAO 15 and DESI DR2), we numerically evaluate this integral and draw the plot which is given in Figure \ref{fig:f8}. We find: $H_{0}t_{0}=0.976$ $\Rightarrow$ $t_{0}=13.53$ Gyr, which is in excellent agreement with recent observational estimates: Planck $2018$: $t_{0}=13.80\pm0.02$ Gyr \cite{Planck2018} and Late-Universe observations (SH0ES): $t_{0}=13.3-13.6$ Gyr \cite{Riess2022}. This agreement indicates that our modified gravity model not only accommodates an accelerated expansion but also reproduces a cosmic age consistent with the current understanding of the Universe’s evolutionary timeline.
\section{Conclusion}\label{sec9}
\hspace{0.5cm} In this work, we have investigated the cosmological dynamics of a Universe governed by the modified gravity theory $f(Q,B)$, where the gravitational Lagrangian is taken as $f(Q,B)= \delta Q^2+\beta B$. To describe the matter content, we employed the Modified Chaplygin Gas (MCG) equation of state $p=A\rho-\frac{B}{\rho^{\alpha}}$, which offers a unified framework for describing the transition from a matter-dominated to a dark energy–dominated universe. Within this framework, we derived an explicit form of the Hubble parameter $H(z)$ as a function of redshift.

Using a robust statistical approach based on MCMC analysis, we constrained the model parameters by employing a comprehensive set of observational data: $46$ Hubble parameter measurements, $15$ BAO points including the recent DESI DR2 BAO sample and the Pantheon+ Type Ia Supernovae compilation. The best-fit values of the free parameters were found to be: $H_0 = 72.2160^{+3.6425}_{-4.4619} \, \text{km/s/Mpc}, \quad A_s = 0.6965^{+0.0817}_{-0.1291}, \quad \alpha = 0.0029^{+0.0225}_{-0.0212}, \quad A = 0.0038^{+0.0714}_{-0.0475}$. These results are consistent with recent observational estimates, including local $H_0$ measurements, and show no tension with the $\Lambda$CDM values within $1–2$$\sigma$ levels.

We then examined the behavior of several cosmological parameters. The deceleration parameter $q(z)$ exhibits a transition from deceleration to acceleration at redshift $z_{tr}\approx0.946$, with a present-day value of $q_0=-0.789$, which align with expectations from recent acceleration data.

The evolution of energy density and pressure further supports the model’s physical viability: the energy density remains positive across all redshifts, while the pressure becomes negative at late times, as expected for dark energy–dominated eras. The EoS parameter $\omega(z)$ begins near zero at early times (indicating matter-like behavior), decreases during intermediate redshifts and asymptotically approaches $\omega_0 \approx -0.691$ at the present epoch. Eventually, it tends toward $\omega = -1$, mimicking the behavior of a cosmological constant.

The model's compliance with energy conditions was also analyzed. We found that the NEC and the DEC are satisfied throughout the cosmic evolution, while the SEC is violated at late times—a typical signature of accelerated expansion driven by dark energy.

We also explored the dynamics in the $\omega-\omega'$ plane. The trajectory starts in the matter-like regime with mild evolution, undergoes a transitional phase marked by a minimum in $\omega'$, and finally settles into a freezing regime consistent with a cosmological constant–like behavior, as the trajectory approaches $(\omega, \omega') = (-1, 0)$.

Finally, we computed the age of the universe in this model using the derived Hubble parameter, obtaining $H_{0}t_{0}=0.976$, which corresponds to $t_0 = 13.53$ Gyr. This result is in excellent agreement with current age estimates from Planck and other late-time cosmological observations.

In summary, the $f(Q,B)$ gravity model with a Modified Chaplygin Gas provides a consistent and observationally viable description of the universe’s accelerated expansion. The model smoothly unifies the early-time matter-dominated era with the current dark energy–dominated epoch and remains compatible with cosmological datasets, thereby offering a consistent framework to explore late-time cosmic acceleration within the context of modified gravity.

\end{document}